\documentstyle[12pt]{article}

\newcommand{\bmat}{\left(\begin{array}}
\newcommand{\emat}{\end{array}\right)}
\def\NPB#1#2#3{Nucl. Phys. B{#1} (19#2) #3}
\def\PLB#1#2#3{Phys. Lett. B{#1} (19#2) #3}

\def\PRD#1#2#3{Phys. Rev. D{#1} (19#2) #3}

\def\MODA#1#2#3{Mod. Phys. Lett.  {#1} (19#2) #3}

\def\yzero{\smash{\hbox{$y\kern-4pt\raise1pt\hbox{${}^\circ$}$}}}

\def\-{\hphantom{-}}

\def\s2{\frac{1}{\sqrt2}}

\def\beq{\begin{equation}}
\def\eeq{\end{equation}}
\def\beqa{\begin{eqnarray}}
\def\eeqa{\end{eqnarray}}
\def\tr{{\rm tr \,}}
\def\Tr{{\rm Tr \,}}
\def\diag{{\rm diag \,}}

\def\IF{\relax{\rm I\kern-.18em F}}
\def\II{\relax{\rm I\kern-.18em I}}
\def\IP{\relax{\rm I\kern-.18em P}}

\def\NN{{\cal N}}

\def\Dsl{\,\raise.15ex\hbox{/}\mkern-13.5mu D} 

\def\IC{\bf C}
\def\IZ{\bf Z}
\def\z2z2{$\IC^3/(\IZ_2\times\IZ_2)$}

\newcommand{\drawsquare}[2]{\hbox{%
\rule{#2pt}{#1pt}\hskip-#2pt
\rule{#1pt}{#2pt}\hskip-#1pt
\rule[#1pt]{#1pt}{#2pt}}\rule[#1pt]{#2pt}{#2pt}\hskip-#2pt
\rule{#2pt}{#1pt}}

\newcommand{\fund}{\raisebox{-.5pt}{\drawsquare{6.5}{0.4}}}
\newcommand{\antifund}{\overline{\fund}}

\topmargin
-1.5cm
\textwidth
15.5cm
\textheight
24cm
\oddsidemargin
0.7cm
\evensidemargin
1.2cm

\begin{document}

\makeatletter
\@addtoreset{equation}{section}
\makeatother
\renewcommand{\theequation}{\thesection.\arabic{equation}}
\pagestyle{empty}
\rightline{IASSNS-HEP-99/94}

\rightline{\tt hep-th/9910155}
\vspace{0.5cm}
\begin{center}
\LARGE{ A New Orientifold of $\IC^2/\IZ_N$ \\
and Six-dimensional RG Fixed Points\\[10mm]}
\large{
Angel~M.~Uranga \footnote{\tt angel.uranga@cern.ch}\\[2mm]}
{\em Theory Division, CERN}\\
{\em CH-1211 Geneva 23, Switzerland} \\[4mm]

\vspace*{2cm}

\small{\bf Abstract} \\[7mm]
\end{center}

\begin{center} \begin{minipage}[h]{14.0cm}
{\small
We discuss the consistency conditions of a novel orientifold projection of
type IIB string theory on $\IC^2/\IZ_N$ singularities, in which one mods
out by the combined action of world-sheet parity and a geometric operation
which exchanges the two complex planes. The field theory on the
world-volume of D5-brane probes defines a family of six-dimensional RG
fixed points, which had been previously constructed using type IIA
configurations of NS-branes and D6-branes in the presence of O6-planes.
Both constructions are related by a T-duality transforming the set of
NS-branes into the $\IC^2/\IZ_N$ singularity. We also construct additional
models, where both the standard and the novel orientifold projections are
imposed. They have an interesting relation with orientifolds of ${\bf
D_K}$ singularities, and provide the T-duals of certain type IIA
configurations containing both O6- and O8-planes.
}

\end{minipage}
\end{center}

\newpage
\setcounter{page}{1}
\pagestyle{plain}
\renewcommand{\thefootnote}{\arabic{footnote}}
\setcounter{footnote}{0}

\section{Introduction}

One of the most interesting quantum field theory lessons that we have
learned from string theory
is the existence of six-dimensional supersymmetric field theories with
non-trivial infrared dynamics \cite{sw,seiberg}. We are now
familiar with the existence of large families of interacting superconformal
field theories with $(0,2)$ and $(0,1)$ supersymmetry. There are two approaches
which have been extensively used to construct these theories. The first
is the study of type IIB D5-branes at A-D-E singularities \cite{dm,jm} and
orientifolds thereof \cite{intri1,bi1,bi2} \footnote{Strings in
orientifold backgrounds have been studied for instance in 
\cite{sag1,dlp,horava,sag2,bs,gp,gjdp}.}. The second is the construction of
type IIA brane configurations (in the spirit of \cite{hw}) of NS-branes,
D6-branes and possibly D8-branes and orientifold planes \cite{bk1,bk2,hz6d}
\footnote{A third approach, the study of F-theory on elliptically
fibered singular Calabi-Yau threefolds \cite{aspmorr} will not be so
relevant for our purposes in the present paper.}

There is a close relation between both constructions. In fact, Type
IIA brane configurations where one of the directions (along which the
D6-branes have finite extent) is compact can be T-dualized to a system of
D5-branes probing orbifold and orientifold singularities (see
{\em e.g.} \cite{kls}). This observation has led to a rich interplay
between both approaches. For example, the construction of superconformal
field theories in the Type
IIB picture in \cite{dm,intri1,bi1,bi2} was a source of information in the
study of the T-dual Type IIA configurations in \cite{bk2,hz6d}. On the
other hand, some configurations in \cite{bk2,hz6d} (those containing
oppositely charged O8-planes) produced new field theories which required
the existence of new orientifolds of $\IC^2/\IZ_N$. These were constructed
in \cite{pu} making use of the information available from the IIA
construction.

In this paper we continue this program by constructing the Type IIB
orientifolds associated to yet another set of field theories constructed
in \cite{bk2,hz6d}. As we discuss below, the new orientifold of
$\IC^2/\IZ_N$ has some unusual and amusing properties \footnote{Several
group theoretical features of the corresponding projection were pointed
out in \cite{bi2}.}.

\medskip

Before continuing with our introduction, it will be  useful to know more
about the relevant IIA brane configurations \cite{bk2,hz6d}. We consider a
set of $N$
NS-branes with worldvolume along 012345, several stacks of D6-branes
(along 0123456) stretched between them, and an O6-plane parallel to the
D6-branes. The direction $x^6$, along which the D6-branes have finite
extent, is taken compactified on a circle. The field theory on the
non-compact part of the D6-brane worldvolume has $D=6$, $\NN=1$
supersymmetry. This is the six-dimensional
version of the four-dimensional models studied in \cite{lll}.

As determined in \cite{ejs} (see also \cite{egkt} for a worldsheet
derivation of this fact) the RR charge of the O6-plane changes sign
whenever it crosses a NS-brane, and the projection it imposes on the
D6-branes changes accordingly \footnote{When $k$ D6-branes sit on top of
an O6$^-$-plane (which carries $-4$ units of D6-brane charge, as counted
in the double cover) their gauge group is projected down to $SO(k)$. When
they sit on top of an O6$^+$ (with charge $+4$) their gauge group is
$USp(k)$.}. Therefore, a consistent  configuration is obtained only for an
even number of NS-branes. If we place $n_i$ D6-branes in the $i^{th}$
interval between NS-branes, the gauge group is of the form
\beqa
SO(n_0)\times USp(n_1)\times\ldots\times SO(n_{N-2})\times USp(n_{N-1})
\label{ft1}
\eeqa
The matter content arises from strings stretching betweeen neighbouring
D6-branes, and is of the form
\beqa
\sum_{i=0}^{N-1} \frac 12 (\fund_i,\fund_{i+1})
\label{ft2}
\eeqa
where the index $i$ is defined mod $N$, and the $\frac 12$ means the
matter arises in half-hypermultiplets, due to the O6-plane projection.
In brane configurations realising six-dimensional field theories, the
NS-branes have the same number of non-compact dimensions as the D6-branes,
so their worldvolume fields are dynamical. In our model, they give rise to
$N-1$ tensor multiplets; an additional tensor multiplet is decoupled
and hence irrelevant.

As discussed in \cite{bk1,bk2,hz6d}, cancellation of charges in the
NS-brane worldvolume imposes a consistency condition on the
configuration, which turns out to be equivalent to the cancellation of
irreducible gauge anomalies \cite{erler} in the six-dimensional field
theory. In our case, these conditions are satisfied for
\beqa
n_{2i}=N+8 \quad ;\quad n_{2i+1} = N
\eeqa
The residual gauge anomaly, as discussed in \cite{bi2}, is cancelled by a
Green-Schwarz mechanism mediated by the tensor multiplets \cite{sagnan}.

As mentioned above, upon T-duality along $x^6$ the NS-branes transform,
roughly speaking, into a $\IC^2/\IZ_N$ singularity \cite{oovafa}, while
the D6-branes become a set of D5-branes sitting at the singularity. The
projection imposed by the O6-plane must transform into an orientifold
projection of $\IC^2/\IZ_N$, such that the field theory (\ref{ft1}),
(\ref{ft2}) arises on the worldvolume of the D5-brane probes. Also, the
$N-1$ tensor multiplets are expected to arise from closed string twisted
sectors.

The orientifold of $\IC^2/\IZ_N$ whose existence is predicted by this
argument has a number of unusual features. For instance, it must exist
only for even $N$. Also, the closed string sector must give rise to one
tensor multiplet per twisted sector. This is in sharp contrast with the
usual orientifold projection considered in \cite{gjdp,bi1,pu}, where
invariant fields are combinations of modes appearing in oppositely twisted
sectors of the type IIB theory. These orientifolds provide one tensor
multiplet and one hypermultiplet per pair of oppositely twisted sectors.
The $N-1$ tensors in our model suggests that the required orientifold
projection is rather different from the familiar one, in that it must map 
each twisted sector to itself.

A second interesting fact is that matter arises in half-hypermultiplets.
Before the orientifolding, the matter in the worldvolume of D5-branes at a
$\IZ_N$ orbifold singularity appears in full hypermultiplets, with the two
`halves' being associated to the two complex planes in $\IC^2/\IZ_N$. In
order to obtain half-hypermultiplets after orientifolding, $\Omega$ must 
be accompanied by a geometric action $\Pi$ which maps the two complex
planes to each other \footnote{Orientifold projections involving
exchange of complex planes have also appeared in \cite{blumen1,blumen2} in
a different kind of models.}.

Finally, the gauge group on the D5-branes suggests the Chan-Paton matrix
associated to the orientifolds projection contains symmetric and
antisymmetric pieces. This contradicts the usual rule that D-branes of the
same dimension suffer orientifold projections of the same kind.

In Section~2 we define the new orientifold projection  $\Omega \Pi$ of
Type IIB on $\IC^2/\IZ_N$, and show it has all the features just
described. In Section~3 we discuss the construction of models where both
the $\Omega \Pi$ and the usual $\Omega$ projections are imposed. 

\section{Orientifold construction}

Here we propose a Type IIB T-dual realization of the field theory in the
previous Section, in terms of D5-branes probing a new kind of orientifold
of $\IC^2/\IZ_N$. As discussed below, we propose that the orientation
reversing element has a geometric action on the $\IC^2/\IZ_N$, $xy=v^N$,
given by $x \leftrightarrow y, v\to-v$. This action is a symmetry only
when $N$ is
even, and involves the exchange of the two complex planes. This  choice
can be heuristically motivated by directly T-dualizing the IIA model
described above. Following \cite{oovafa}, upon T-duality the set of $N$
NS-branes becomes a $N$-centered Taub-NUT space, which can be described
as $xy=v^N$ in suitable complex coordinates, while the D6-branes
transform into D5-brane probes. The directions 89 can be identified with
$v$. The directions 7 and 6$'$ (the T-dual of 6, with asymptotic radius
$R_{6'}$) are related to $x$ and $y$ in a complicated manner. Roughly
speaking, for large $x$ and fixed $y$ one has $x\sim
e^{i(x^7+ix^{6'})/R_{6'}}$ and for large $y$ at fixed $x$ one has
$y\sim e^{-(x^7+ix^{6'})/R_{6'}}$. The T-dual of the O6-plane action is
expected to flip the signs of the coordinates 6$'$789, which is certainly
the case for the action proposed above.

This argument should be considered a heuristic motivation. Stronger checks
will arise from the detailed discussion of the model. We therefore turn to
constructing the new orientifold of $\IC^2/\IZ_N$ and to showing that the
world-volume field theory on D5-brane probes is the one described in
Section~1. Let us consider type IIB theory on $\IC^2/\IZ_N$, modded out by
$\Omega \Pi$, where $\Omega$ is world-sheet parity and $\Pi$ acts as
\beq
\Pi: \;\; z_1\to z_2 \quad, \quad z_2 \to -z_1
\eeq
with $z_1$, $z_2$ parametrizing the two complex planes in $\IC^2/\IZ_N$.
Consistency requires $(\Omega\Pi)^2$ to belong to the orbifold group
$\IZ_N$. Since it acts as $z_i\to -z_i$, this is the case when $N$ is even,
$N=2K$. This condition, which agrees with a constraint found in the T-dual
type IIA brane configuration, will be reobtained below from a different
point of view. The orientifold defined above preserves $\NN=1$
supersymmetry in six dimensions.

\subsection{Closed string spectrum}

As already mentioned, a first peculiar feature of this orientifold appears
in the computation of the closed string twisted sectors. In order to
appreciate this, it will be convenient to compare the situation with the
usual $\Omega$ projection.

In type IIB theory on $\IC^2/\IZ_N$, the left and right moving states are
labeled, in the bosonic description, by a vector $r_k=(r_1,r_2,r_3,r_4)$
belonging to the $SO(8)$ weight lattice shifted by the twist vector
$v_k=(k/N,-k/N,0,0)$, and constrained by $\sum_{i=1}^4 r_i= 1\ {\rm
mod} \ 2$. At the massless level, these states are
\beqa
NS: & \quad & |\Phi_k^{(1)} \rangle = |{\small -1+\frac kN, -\frac
kN, 0, 0 }\rangle \nonumber \\
    & \quad & |\Phi_k^{(2)} \rangle = |{\small \frac kN, 1-\frac
kN, 0, 0 }\rangle \nonumber \\
R: & \quad & |\Psi_k^{(1)} \rangle = |{\small -\frac 12 + \frac kN, \frac
12 -\frac kN, \frac 12, -\frac 12 }\rangle \nonumber \\
   & \quad & |\Psi_k^{(2)} \rangle = |{\small -\frac 12 + \frac kN, \frac
12 -\frac kN, -\frac 12, \frac 12}\rangle
\eeqa
Their $\theta$ eigenvalue is given by $r_k\cdot v_k=-{k\over N}(1-2k/N)$.
The NS-NS and R-R $\IZ_N$ invariant states are \footnote{The NS-R and R-NS
states can be determined from these by supersymmetry.}
\beqa
\begin{array}{cccc}
{\rm NS-NS}: & \ |\Phi_k^{(i)}\rangle_L \otimes |\Phi_{N-k}^{(j)}\rangle_R
& \ 4(1,1) & \ i,j=1,2 \\
{\rm R-R}: & \ |\Psi_k^{(i)}\rangle_L \otimes |\Psi_{N-k}^{(j)}\rangle_R &
\ (1,1) + (1,3) & \ k=1,\ldots,N-1
\end{array}
\eeqa
The second column gives the representation under the spacetime little
group $SU(2)\times SU(2)$. We obtain one hyper and one tensor multiplet of
$D=6$ $\NN=1$ supersymmetry per twisted sector.

In the usual $\Omega$ orientifolds studied in the literature \cite{gjdp},
and for generic twisted sectors ($k\neq N/2$), the states surviving the
projection are
\beqa
\begin{array}{ccc}
{\rm NS-NS}: & \ |\Phi_k^{(i)}\rangle_L \otimes |\Phi_{N-k}^{(j)}\rangle_R
\; + \; |\Phi_{N-k}^{(j)}\rangle_L \otimes |\Phi_k^{(i)}\rangle_R & \
4(1,1) \\
{\rm R-R}: & \ |\Psi_k^{(i)}\rangle_L \otimes |\Psi_{N-k}^{(j)}\rangle_R
\; - \; |\Psi_{N-k}^{(j)}\rangle_L \otimes |\Psi_k^{(i)}\rangle_R &
\ (1,1)+(1,3) \\
\end{array}
\eeqa
Notice that invariant combinations are a mixture of states appearing in
oppositely twisted sectors of the type IIB theory. The orientifolded model
produces one hyper and one tensor multiplet per such pair of oppositely
twisted sectors. 

For the $k=N/2$ twisted sector, which does not mix with any other, the
action of $\Omega$ may be defined with an additional $(-1)$ sign
\cite{polchinski}. This $\IZ_2$ choice determines the type of multiplet
surviving the projection. One gets a tensor multiplet or a hypermultiplet
if the additional sign is present or not. As discussed in
\cite{polchinski}, consistent coupling between closed and open strings 
implies the following constraint on the D-brane Chan-Paton matrices
\beq
\gamma_{\theta^{N/2}}=\mp \gamma_{\Omega} \gamma_{\theta^{N/2}}^T
\gamma_{\Omega}^{-1}
\label{polchirel}
\eeq
where the positive sign is taken when $\Omega$ includes the additional
$(-1)$ sign, and the negative sign is taken otherwise. The two
possibilities correspond to D-brane gauge bundles with or without vector
structure \cite{blpssw}.

\medskip

Let us now turn to the $\Omega\Pi$ orientifold. 
It is easy to realize that $\Omega\Pi$ invariant states do not require
mixing different twisted sectors of the type IIB theory. Because of that,
for every sector there is {\em a priori} a $\IZ_2$ ambiguity in defining
the orientifold action. Actually, below we will show that the only consistent
possibility is to include the additional $(-1)$ sign in the action of
$\Omega\Pi$. Assuming momentarily this claim, the resulting invariant
states are
\beqa
\begin{array}{ccc}
{\rm NS-NS:} & \ |\Phi_k^{(1)}\rangle_L \otimes |\Phi_{N-k}^{(2)}\rangle_R
- |\Phi_k^{(2)}\rangle_L \otimes |\Phi_{N-k}^{(1)}\rangle_R & \ (1,1) \\
{\rm R-R}: & \ |\Psi_k^{(1)}\rangle_L \otimes |\Psi_{N-k}^{(1)}\rangle_R &
\\
    & \ |\Psi_k^{(2)}\rangle_L \otimes |\Psi_{N-k}^{(2)}\rangle_R &
\ (1,3)\\
    & \ |\Psi_k^{(1)}\rangle_L \otimes |\Psi_{N-k}^{(2)}\rangle_R +
      |\Psi_k^{(2)}\rangle_L \otimes |\Psi_{N-k}^{(1)}\rangle_R & \\
\end{array}
\eeqa
We obtain one tensor multiplet per twisted sector. We now show that this
choice is the only consistent one, by using the constraints coming from
consistent coupling of open and closed strings. Consider a set of
D5-branes at the $\IC^2/\IZ_N$ singularity, before the orientifold
projection is imposed \footnote{The $\Omega\Pi$ orientifold
does not allow for the introduction of D9-branes. In Section~3 D9-branes
may appear in some models where both the $\Omega\Pi$ and the usual $\Omega$
projections are imposed. The consistecy conditions for D9-branes in those
cases can derived by simple modifications in our arguments below.}. Their
Chan-Paton matrix $\gamma_{\theta,5}$ has $N$ different eigenvalues
$s_i$ with multiplicities $n_i$. It can be shown that
$\gamma_{\Omega\Pi,5}$ is block-diagonal (with $n_i\times n_i$ blocks) in
this basis, and therefore $\gamma_{\theta^k,5}$ and $\gamma_{\Omega\Pi,5}$
commute. The fact that $\gamma_{\Omega\Pi}$ acts diagonally on the index
$i$ can be derived {\em e.g.} by looking at the Chern-Simons couplings of
open string states and closed string RR modes in the $\IZ_N$ orbifold
\cite{dm}
\beqa
\int \tr(\gamma_{\theta^k,5}\lambda_i) C_k \wedge e^{F_i}
\eeqa
where $C_k$ is a formal sum of the twisted RR forms, and $\lambda_i$
selects the entries of the Chan-Paton matrix corresponding to the $i^{th}$
set of D-branes. Since $\Omega\Pi$ is a symmetry of the orbifold theory,
the coupling must be invariant. Recalling that $\Omega\Pi$ acts diagonally
in $k$, it follows that its action on the open string sector is diagonal
in $i$ \footnote{In other words, $i$ labels the different kinds of 
fractional branes \cite{fract}, which can be understood as
higher-dimensional branes wrapped on the collapsed two-cycles of the
orbifold singularity. Since the orientifold acts diagonally on the cycles,
so it does on the wrapped branes.}.

The argument in \cite{polchinski} now allows us to use this information on
the open string sector to constrain the closed string sector. Since
$\gamma_{\Omega\Pi}$ commutes with $\gamma_{\theta^k}$, we have
\beqa
\gamma_{\theta^k,5}= +\gamma_{\Omega\Pi,5} \gamma_{\theta^k,5}^T
\gamma_{\Omega \Pi,5}^{-1}
\eeqa
Comparing with (\ref{polchirel}), we learn that the projection $\Omega\Pi$
in the closed string sector must include the additional $(-1)$ sign, as
claimed above. This fixes the $\IZ_2$ ambiguity in all twisted sectors.

The bottomline of this argument is that the closed string spectrum of the
$\Omega\Pi$ orientifold of $\IC^2/\IZ_N$ produces $N-1$ tensor multiplets.

\subsection{Open string spectrum}

We now turn to the open string spectrum and further consistency
conditions on $\gamma_{\Omega\Pi,5}$. It is convenient to introduce a
concrete expression for $\gamma_{\theta,5}$, which we take \footnote{In
the $\Omega$ orientifolds, a different model would be obtained for
$\gamma_{\theta^N,5}=- 1$. It is easy to show that they are identical in
the $\Omega\Pi$ orientifold.}
\beqa
\gamma_{\theta,5}=\diag(1_{n_0},e^{2\pi i\frac 1N}1_{n_1},\ldots,
e^{2\pi i\frac{N-1}{N}} 1_{n_{N-1}})
\eeqa

An interesting constraint on $\gamma_{\Omega\Pi,5}$ can be derived by
requiring $(\Omega\Pi)^2=\theta^{N/2}$ in the open string sector. The
action of $\Omega\Pi$ on a state $|\psi,ij\rangle$, corresponding to an
open string stretching between the $i^{th}$ and $j^{th}$ D-brane, is
\beqa
\Omega\Pi: |\psi,ij\rangle \to (\gamma_{\Omega\Pi})_{ii'} |\Omega\Pi\cdot
\psi, \ j'i'\rangle (\gamma_{\Omega\Pi}^{-1})_{j'j}
\eeqa
We are interested in comparing the actions of $(\Omega\Pi)^2$ and
$\theta^{N/2}$, given by
\beqa
& (\Omega\Pi)^2: & |\psi,ij\rangle \to (\gamma_{\Omega\Pi,5}
(\gamma_{\Omega\Pi,5}^T)^{-1})_{ii'} |\Omega\Pi\cdot \psi, i'j'\rangle
(\gamma_{\Omega\Pi,5}^{T} \gamma_{\Omega\Pi,5}^{-1} )_{j'j} \nonumber\\
& \theta^{N/2}: & |\psi,ij\rangle \to (\gamma_{\theta^{N/2},5})_{ii'}
|\theta^{N/2} \cdot \psi, i'j'\rangle (\gamma_{\theta^{N/2},5}^{-1})_{j'j}
\label{constrsymm}
\eeqa
Let us split the set of D-branes into two types, denoted `even' and `odd',
according to whether their $\gamma_{\theta^{N/2}}$ eigenvalue is $+1$ or
$-1$. In the even-even and odd-odd sectors, $\theta^{N/2}$ acts as $+1$,
and (\ref{constrsymm}) imposes $\gamma_{\Omega\Pi}=\pm\gamma_{\Omega\Pi}^T$,
with independent choices of sign for `even' and `odd' D-branes. In the
even-odd and odd-even sectors, $\theta^{N/2}$ acts as $-1$ and
(\ref{constrsymm}) implies the symmetry of $\gamma_{\Omega\Pi}$ must be
opposite for `even' and `odd' branes. Without loss of generality we get
the conditions
\beqa
& \gamma_{\Omega\Pi,5}=+\gamma_{\Omega\Pi,5}^T & \quad {\rm for \;\;
even \;\; branes} \nonumber \\
& \gamma_{\Omega\Pi.5}=-\gamma_{\Omega\Pi,5}^T & \quad {\rm for \;\;
odd\;\; branes}
\eeqa
An appropriate choice of $\gamma_{\Omega\Pi,5}$ is
\beqa
\gamma_{\Omega\Pi,5}=\diag(1_{n_0},\epsilon_{n_1},\ldots, 1_{n_{N-2}},
\epsilon_{n_{N-1}})
\label{alfin}
\eeqa
Notice that this type of constraint is satisfied only for $N$
even. We also would like to point out that the different symmetry of
$\gamma_{\Omega\Pi,5}$ for `even' and `odd' branes is related by T-duality
to the change of sign of the O6-plane whenever it crosses a NS-brane. The
$\Omega\Pi $ orientifold hence provides a geometrical description of such
process in a IIB T-dual realisation. It would be interesting to compare it
with the geometries proposed in \cite{pru}.

\medskip

This completes the discussion on consistency conditions. Let us turn to
computing the massless open string spectrum. The projection on the
Chan-Paton factors for gauge bosons is
\beqa
\lambda & = & \gamma_{\theta,5}\lambda \gamma_{\theta,5}^{-1} \nonumber \\
\lambda & = & - \gamma_{\Omega\Pi,5} \lambda^T \gamma_{\Omega\Pi,5}^{-1}
\eeqa
and leads to a gauge group
\beqa
SO(n_0)\times USp(n_1)\times\ldots\times SO(n_{N-2})\times USp(n_{N-1})
\label{group1}
\eeqa
The projection on the matter Chan-Paton factors is
\beqa
\begin{array}{lll}
Z_1  =  e^{2\pi i/N}\ \gamma_{\theta,5} Z_1 \gamma_{\theta,5}^{-1} & \quad
& Z_2  =  e^{-2\pi i/N}\ \gamma_{\theta,5} Z_2 \gamma_{\theta,5}^{-1} \\
Z_2 = \gamma_{\Omega\Pi,5}\ Z_1^T\ \gamma_{\Omega\Pi,5}^{-1} & \quad
& Z_1 = - \gamma_{\Omega\Pi,5}\ Z_2^T\ \gamma_{\Omega\Pi,5}^{-1}
\end{array}
\eeqa
We obtain the following hypermultiplet matter content
\beqa
\sum_{i=0}^{N-1}\  \frac 12 (\fund_i,\fund_{i+1})
\label{matt1}
\eeqa
Notice that the action of $\Pi$ as exchange of the two complex planes is
essential in obtaining half (rather than full) hypermultiplets.

Thus, the spectrum in the closed and open string sectors agrees with the
field theory constructed in Section~1 from type IIA brane configurations
of NS- and D6-branes in the presence of an O6-plane.

This field theory is potentially anomalous. We now show that, in
analogy with the models in \cite{bi1,bi2,pu}, the irreducible gauge
anomaly
vanishes once tadpole cancellation conditions are imposed. The reader not
interested in these details in encouraged to skip the computation. The
tadpoles can be obtained using the general techniques in \cite{gp,gjdp},
and we only stress the differences between the $\Omega$ and $\Omega \Pi$
projections. Since the cylinder diagrams do not involve crosscaps, their
contribution to the tadpoles is the familiar one
\beqa
{\cal C}=\sum_{k=1}^{N-1} 4\sin^2(\frac{\pi k}{N}) (\Tr
\gamma_{\theta^k,5})^2
\eeqa
where the untwisted tadpole $k=0$ vanishes in the non-compact limit and is
therefore ignored. The tadpoles from the M\"obius strip diagrams are
quite similar to those in \cite{gjdp}. The only difference is that the
eigenvalues of the twists $\theta^k\Pi$ that act along with $\Omega$ are
$e^{\pm 2\pi i/4}$ (and independent of $k$) rather than $e^{\pm 2\pi
ik/N}$, and this modifies the trigonometric coefficient of the tadpole.
Concretely we have
\beqa
{\cal M} = -16 \sum_{k=0}^{N-1} 4 \sin^2 \frac{\pi}{4} \Tr
(\gamma_{\theta^k\Omega\Pi}^T \gamma_{\theta^k\Omega\Pi}^{-1}) = -32 N \Tr
\gamma_{\theta^{N/2}}
\eeqa
where we have used the property $\Tr(\gamma_{\theta^k\Omega\Pi}^T
\gamma_{\theta^k\Omega\Pi}^{-1})=\Tr\gamma_{\theta^{N/2}}$. The Klein bottle
amplitude ${\cal K}(\theta^n,\theta^k)$ is evaluated by tracing over the
$\theta^n$-twisted closed string spectrum with an insertion of
$\theta^k\Omega\Pi$. Since $\Omega\Pi$ acts diagonally in the twisted
sector index, all values of $n$ contribute \footnote{Recall that in the
$\Omega$ orientifolds only the pieces with $n=0,N/2$ give a net
contribution. For other
twists, the contributions from symmetric and antisymmetric combinations of
$\theta^n$ and $\theta^{-n}$ twisted sectors cancel in the trace.}. We must
take into account that, as in the M\"obius strip computation, the
eigenvalues for the twist $\theta^k\Pi$ are $e^{\pm 2\pi i/4}$, and that
for $n=0$ zero mode factors from momentum states should be included.
The different contributions ${\cal T}(\theta^n,\theta^k)$ to the tadpoles
add up to 
\beqa
{\cal K} & = & \sum_{k=0}^{N-1} {\cal T}(1,\theta^k) + \sum_{n=1}^{N-1}
\sum_{k=0}^{N-1} {\cal T}(\theta^n,\theta^k) = \nonumber \\
& & = 64 \sum_{k=0}^{N-1} \frac{4 \sin^2 \frac{2\pi}{4}}{4\sin^2 \frac{\pi}{4}
4\sin^2\frac{3\pi}{4}} + 64 \sum_{n=1}^{N-1} \sum_{k=0}^{N-1} 1 \; = \;
64 N^2
\eeqa
The complete expression ${\cal C}+{\cal M}+{\cal K}$ can be factorized as
\beqa
\sum_{k=1}^{N-1} \frac{1}{\textstyle 4 \sin^2 \frac{\pi k}{N}}\ \left(\;
4\sin^2\frac{\pi k}{N} \Tr \gamma_{\theta^k,5} - 16 N \delta_{k,N/2}\;
\right)^2 = 0
\eeqa
and leads to the constraint
\beqa
4\sin^2\frac{\pi k}{N} \Tr \gamma_{\theta^k,5} - 16 N \delta_{k,N/2}\; = 0
\eeqa
The tadpole cancellation conditions can be expressed in terms of the
integers $n_r$, yielding the condition
\beqa
-2 n_r + 16(-1)^{r} + n_{r-1} + n_{r+1} = 0
\eeqa
This is the condition of cancellation of the irreducible anomaly in the
field theory (\ref{group1}), (\ref{matt1}). Their solution is $n_{2i}=N+8$,
$n_{2i+1}=N$. This solution reproduces the result based on RR
charge conservation in the type IIA brane configuration discussed in
the introduction. The factorization  of the remaining anomalies and their
cancellation by a Green-Schwarz  mechanism involving the $N-1$ tensor
multiplets \cite{sagnan} has been discussed in \cite{bi2}, and we will not
repeat it here.

\section{Theories with the ${\bf \Omega\Pi}$ and ${\bf \Omega}$
projections}

In this section we comment on the consistency conditions in type IIB
orientifolds where both the usual $\Omega$ and the new $\Omega\Pi$
projections are imposed. As in the construction of the $\Omega\Pi$
orientifold, the T-dual type IIA brane configurations will provide a
useful guideline. We are by now familiar with the fact that the
$\Omega\Pi$ projection corresponds to the presence of an O6-plane in the
T-dual Type IIA model. Similarly, the $\Omega$ projection corresponds to
the presence of two O8-planes (along 012345789) in the IIA configuration.
Hence the IIA  brane configurations for the models in this section include
orientifold projections by both O6- and O8-planes.  The general features
of these configurations have been discussed in \cite{hz6d}, and in
\cite{hzdk} in more detail.

We consider type IIB on $\IC^2/\IZ_N$, with $N=2K$, modded out by the
orientifold projections $\Omega\Pi$ and $\Omega$. A first observation
is that closure requires the introduction of the element $\Pi$ in the
orbifold group. The generators $\theta$ and $\Pi$ satisfy the properties
\beqa
\theta^N=1 \quad ; \quad \Pi^2=\theta^{K} \quad ;\quad \Pi \ \theta^n = 
\theta^{-n} \ \Pi
\eeqa
These relations define the non-abelian discrete group ${\bf D_{K}}$. Hence
we are dealing with orientifolds of  $\IC^2/D_{K}$. How does this arise in
the IIA brane configuration?

\subsection{The IIA brane configurations}

The answer goes as follows. In a type IIA brane configuration with O6-
and O8-planes, the orientifolds impose the projections
$\Omega_{O6}\equiv\Omega (-1)^{F_L}R_{7}R_8R_9$ and
$\Omega_{O8}\equiv\Omega R_6$, respectively, where $R_i$ acts as $x^i\to
-x^i$. By closure, the configuration is also modded out by the
{\em orbifold} element ${\cal R}=(-1)^{F_L}R_{6}R_7R_8R_9$. 

It is illustrative to consider momentarily the configuration modded out
only by ${\cal R}$, without any orientifold projection. Such brane
configurations were studied in \cite{kapustin} and shown to reproduce
field theories with gauge groups and matter contents defined by ${\bf D_K}$
quiver diagramas \cite{jm}. Furthermore, these configurations have been 
argued to be T-dual to systems of D5-branes at $\IC^2/D_K$ singularities
\cite{hzdk}. The type IIA brane configurations we are actually interested
in contain an additional orientifold projection, say, by $\Omega_{O6}$.
Hence their type IIB T-duals should correspond to orientifolds of
$\IC^2/D_K$ singularities, as found above.

Let us consider the IIA brane configurations is some more detail. Since
they contain an O6-plane which flips charge whenever it crosses a NS-brane, 
$N$ is constrained to be even, $N=2K$. Another consistency condition is
that no NS-brane can intersect the O8-planes, since the O6-plane crossing
them would not respect the $\IZ_2$ symmetry imposed by the O8-plane. For a
fixed $K$, in order to specify the model completely one has to make the
following choices. First, the charge of the two O8-planes \footnote{Our
convention is that an O8$^+$-plane has $+16$ units of D8-brane charge, and
projects the gauge group of $k$ coincident D6-branes down to $SO(k)$, and
that an O8$^-$-plane, with $-16$ units of D8-brane charge, produces a
gauge group $USp(k)$ on $k$ D6-branes.}, which can be positive for one and
negative for the other -- in which case their RR charge cancels and no
D8-branes are required -- or negative for both --in which case RR charge
cancellation requires 32 D8-branes, as counted in the covering space; in
this case one also has to choose the distribution of the D8-branes in the
different intervals, in a way consistent with the orientifold symmetries
--. Second, the charge assignment for the different pieces of O6-plane in
the intervals. Finally, the number of D6-branes $n_i$ at each interval $i$
is determined by the conditions or RR charge conservation \cite{hz6d,
bk2}. The solution to these conditions is unique up to an overall addition
of an equal number of D6-branes in all intervals. 

To make the discussion a bit more concrete, consider the case of even $K$.
There are two possible patterns for the gauge group
\beqa
& G_0 \times USp(n_1)\times SO(n_2) \times \ldots \times USp(n_{K-1})
\times G_{K} & \nonumber \\
& G_0 \times SO(n_1)\times USp(n_2) \times \ldots \times SO(n_{K-1})
\times G_{K} & 
\label{group2}
\eeqa
The fist (resp. second) possibility corresponds to the case when the
O8-planes are intersected by O6$^-$- (resp. O6$^+$)- planes. The nature of
the factors $G_0$ and $G_K$ is more model-dependent and will be discussed
below. When $K$ is odd, the general structure of the gauge group is 
\beqa
& G_0 \times USp(n_1)\times SO(n_2) \times \ldots \times
USp(N_{K-2})\times SO(n_{K-1}) \times G_{K} &
\label{group3}
\eeqa
In any case, the  general structure of the matter content is given by
\beqa
\frac 12 (R_0,\fund_1) + \sum_{i=1}^{K-2} \ \frac 12(\fund_i,\fund_{i+1})
+ \frac 12 (\fund_{K-1},R_K)
\label{matt2}
\eeqa
where $R_0$ and $R_K$ are discussed below. If D8-branes are present,
suitable flavours in fundamental representations must be added
\cite{hzdk}.

Let us now specify the structure of the `end' sectors in the above
spectra, where a certain number $n$ of D6-branes suffers the projection by
an O8-plane and an O6-plane. There are four different cases to be
considered, depending on the charges of the orientifold planes, and
they lead to different group factors and matter contents \cite{hzdk}.

{\bf i)} When the projection is imposed by an O8$^-$-plane and  an
O6$^-$-plane,  the corresponding gauge factor is $SU(n/2)$ and the
representation $R_0$ or $R_K$ in (\ref{matt2}) is $\fund +\antifund$.

{\bf ii)} When there is an O8$^+$-plane and an O6$^+$-plane, we obtain
the same answer: the gauge factor is $SU(n/2)$ and  the representation
$R_0$ or $R_K$ is $\fund +\antifund$.

{\bf iii)} For an O8$^-$-plane and an O6$^+$ plane, the gauge factor is
$USp(n) \times USp(n')$ and the representation $R_0$ or $R_{K}$ is
$(\fund,1) + (1,\fund)$.

{\bf iv)} For an O8$^+$-plane and an O6$^-$ plane, the gauge factor is
$SO(n) \times SO(n')$ and the representation $R_0$ or $R_{K}$ is
$(\fund,1) + (1,\fund)$.

Concerning the NS-brane world-volume fields, we obtain $K$ tensor
multiplets from the $K$ NS-branes in the configuration. There are also
twisted fields living at the fixed points of $(-1)^{F_L} R_6 R_7 R_8 R_9$.
For `end' sectors of the type {\bf i)} or {\bf ii)} each orbifold plane gives 
rise to a hypermultiplet. For `end' sectors of the type {\bf iii)} or 
{\bf iv)}, each produces one tensor multiplet.

In the resulting field theory, the ranks of the gauge groups are
determined by imposing the charge cancellation conditions in the IIA brane
configuration. These include the the charges with respect to the twisted
fields of the orbifold plane. In the cases without D8-branes, on which we
center henceforth for the sake of clarity, this last condition implies
$n=n'$ in the cases {\bf iii)} and {\bf iv)}.

In the following subsection we show how these rules arise in the
construction of the Type IIB orientifolds,

\subsection{The orientifold construction}

Even though as explained above we are dealing with an orientifold of a
${\bf D_K}$ singularity, it will be convenient to continue discussing in
terms of a $\IZ_N$ singularity modded out by the two orientifold projections 
$\Omega \Pi$ and $\Omega$. This is useful since it allows to benefit from
the known results about models with either of the projections, taken from 
Section~2 for the $\Omega \Pi$ and from \cite{bi1,pu} for the $\Omega$ 
projections. There are nevertheless some important instances where the
additional elements $\theta^k\Pi$ in the orbifold group play a role. 

One of these situations is the computation of the closed string spectrum.
The model contains a twisted sector for each one of the $K+2$ non-trivial
conjugacy classes of the orbifold group ${\bf D_K}$. These classes are
\beqa
{\cal C}_n & = & \{ \theta^n,\theta^{-n} \} \quad n=1,\ldots, K-1 \quad,
\quad {\cal C}_{K}  =  \{ \theta^K \} \nonumber \\
{\cal C}_{even} & = & \{ \Pi, \theta^2\Pi,\ldots,\theta^{2K-2}\Pi \}
\quad, \quad {\cal C}_{odd} = \{ \theta\Pi, \ldots, \theta^{2K-1}\Pi \} 
\eeqa
We see that the new elements in the orbifold group generate new twisted
sectors, but also introduce identifications in the twisted sectors of
$\IZ_N$. In type IIB theory, each sector produces a hyper and a tensor
multiplet. In the orientifolded theory, the action of $\Omega$ and
$\Omega\Pi$ on the classes ${\cal C}_{n}$, $n=1,\ldots, K$ selects the
tensor multiplet, while the action on the remaining classes may select the
tensor or hypermultiplet. The relevant cases are discussed in the examples
below. Notice that the action of $\Omega$ on the $\theta^{N/2}$-twisted
sector corresponds to a projection with vector structure.

Another consequence of the new elements in the orbifold group is
that the D5- and D9-branes in the model generate new disk tadpoles
twisted by these elements $\theta^k\Pi$. These do not have a
corresponding crosscap diagram, so cancellation of these tadpoles requires 
conditions of the type
\beqa
\Tr \gamma_{\theta^k\Pi,9} - 2\Tr \gamma_{\theta^k\Pi,5} = 0
\eeqa
where the factor of two arises form momentum zero modes, and the first
term is required only in cases with D9-branes. This condition is the IIB
counterpart of the cancellation of charge under the twisted fields of the
orbifold in the IIA brane configuration. We will make sure this condition
is satisfied for the matrices in our models.

\medskip

Let us turn to the explicit construction.
Intead of being completely general, it will be illustrative to consider
two examples, which include all the different `end' sectors discussed in
section 3.1. Let us consider $\IC^2/\IZ_{N}$ with $N=2K$ and $K$ odd, and
mod it by $\Omega\Pi$ and $\alpha\Omega$, with $\alpha^2=\theta$. As
explained in \cite{pu}, the presence of $\alpha$ in the $\Omega$
projection implies the model contains no D9-branes, a conveniently simple
case. Let us consider the Chan-Paton matrices
\beqa
\gamma_{\theta,5} & = &\diag(1_{n_0},e^{2\pi i\frac{1}{N}} 1_{n_1}, \ldots,
e^{2\pi i\frac{K-1}{N}} 1_{n_{K-1}}, e^{2\pi i\frac{K}{N}} 1_{n_K},
e^{2\pi i\frac{K+1}{N}} 1_{n_{K-1}}, \ldots, e^{2\pi i\frac{N-1}{N}
1_{n_1}}) \nonumber \\
\gamma_{\Omega\Pi,5} & = & \diag(\epsilon_{n_0}, 1_{n_1}, \ldots,
\epsilon_{n_{K-1}}, 1_{n_K},\epsilon_{n_{K-1}}, \ldots, 1_{n_1}) \nonumber
\\
\gamma_{\alpha\Omega} & = & {\small \pmatrix{
1_{n_0}  &  &  &  &  &  &  &  \cr
  &  &  &  &  &  &  & \alpha 1_{n_1} \cr
  &  &  &  &  &  & \cdots &  \cr
  &  &  &  &  & \alpha^{P-1}1_{n_{P-1}} &  &  \cr
  &  &  &  & \alpha^P \varepsilon_{n_P}  &  &  &  \cr
  &  &  & \alpha^{-(P-1)} 1_{n_{P-1}} &  &  &  &  \cr
  &  & \cdots &  &  &  &  &  \cr
  & \alpha^{-1} 1_{n_1} & & & & & & \cr }} 
\label{matrix1} 
\eeqa 
Notice that the matrices $\gamma_{\theta^k\Pi \alpha}= \gamma_{\alpha
\theta^k\Omega} \gamma_{\Pi\Omega}$ are traceless, so the additional disk
tadpoles mentioned above vanish. The open string spectrum one obtains from
the projection using the matrices above is 
\beqa 
& U(n_0/2)\times SO(n_1)\times \ldots \times USp(n_{K-1})\times U(n_K/2) &
\nonumber \\ &
\frac 12 [(\fund_0,\fund_1)+(\antifund_0,\fund_1)] + \sum_{i=1}^{K-2}
\frac 12 (\fund_i,\fund_i) + \frac 12 [(\fund_{K-1},\fund_K) +
(\fund_{K-1},\antifund_K)] & 
\label{spec1} 
\eeqa 
The closed string spectrum contains $K$ tensor multiplets and two
hypermultiplets.  

This model has a clear type IIA T-dual configuration. It
contains $N$ NS branes, two O8-planes (not intersected by any NS-brane)
and one O6-plane. The $\alpha\Omega$ projection above corresponds
to the case where the O8-planes are oppositely charged \cite{pu}. The
choice of Chan-Paton matrices $\gamma_{\Omega\Pi}$,
$\gamma_{\alpha\Omega}$ above specifies that the O8$^+$ is
intersected by an O6$^+$, and the O8$^-$ by an O6$^-$. Using the rules of
section 3.1 we obtain a gauge group and matter content in agreement with
(\ref{spec1}).

Let us turn to the discussion of cancellation of tadpoles in the type IIB
side. Instead of performing it in the usual way, by factorizing Klein
bottle, M\"obius strip and cylinder diagrams, one can take advantage of
knowing directly the contributions of the $\alpha\Omega$ \cite{pu} and
$\Omega\Pi$ crosscaps (Section 2 \footnote{Actually, the slightly
different form of $\gamma_{\Omega\Pi}$ in this section introduces a $-1$
sign in the crosscap computed in Section~1. Also notice that the crosscap
in \cite{pu} should be multiplied by 4, since our models are
six-dimensional.}). The tadpole condition reads
\beqa
4\sin^2 (\frac{\pi k}{N}) \Tr \gamma_{\theta^{k},5} - 32 \delta_{k,1\;
{\rm mod}\; 2} + 16 N \delta_{k,N/2} = 0
\label{tadpole1}
\eeqa
This expression agrees with that obtained performing the standard
computation. This condition can be recast in terms of the integers $n_r$,
giving
\beqa
-2 n_r+ n_{r-1}+ n_{r+1} +16\delta_{r,0} -16\delta_{r,K} -16(-1)^r = 0
\eeqa
These conditions correspond precisely to the constraints from cancellation
of the irreducible anomaly, and to the conditions of RR charge
conservation in the T-dual type IIA brane configuration. The residual
gauge anomaly is factorized and cancelled as discussed in \cite{hz6d,
hzdk} by exchange of the closed string tensor multiplets \cite{sagnan}.
Also, the $U(1)$ anomalies are cancelled by hypermultiplet exchange.

\medskip

The second example we consider is a slight modification of the previous
model. Let us again consider $\IC^2/\IZ_N$, with $N=2K$ and $K$ odd,
modded out by $\alpha\Omega$ and $\Omega \Pi$. Let us choose the
Chan-Paton embedding of $\gamma_{\theta,5}$ and $\gamma_{\alpha\Omega,5}$
to be exactly as in (\ref{matrix1}), but let us take
\beqa
\gamma_{\Omega\Pi,5} = \diag (\sigma_1 \otimes 1_{n_0/2},
\epsilon_{n_1},\ldots, 1_{n_{K-1}}, i\sigma_2\otimes 1_{n_K/2},
1_{n_{K-1}}, \ldots, \epsilon_{n_{1}})
\eeqa
Notice that $\gamma_{\Omega\Pi,5}$ by itself is equivalent to
(\ref{alfin}),
despite the modification of the blocks at the positions $0$ and $K$. The
modification is required so that the matrices 
$\gamma_{\theta^k\Pi\alpha}$ are traceless, as required for consistency.
The final spectrum of this model is
\beqa
& [SO(n_0/2)\times SO(n_0/2)']\times SO(n_1)\times \ldots \times
USp(n_{K-1})\times [USp(n_K/2)\times USp(n_K/2)'] &
\nonumber \\
& \frac 12 (\fund_0,\fund_1)+\frac 12(\fund_0',\fund_1) + \sum_{i=1}^{K-2}
\frac 12 (\fund_i,\fund_i) + \frac 12 (\fund_{K-1},\fund_K) + \frac 12
(\fund_{K-1},\fund_K')] \quad &
\label{spec2}
\eeqa
The closed string spectrum gives $K+2$ tensor multiplets.

The interpretation in terms of the T-dual type IIA brane configuration is
identical to the one we had for the previous model, save for a flip in the
sign of the O6-plane. Hence in this model the O8$^+$-plane is intersected
by an O6$^-$-plane, and the O8$^-$-plane is intersected by an
O6$^+$-plane. The spectrum (\ref{spec2}) agrees with the rules given in
section 3.1, in particular with the `end' sectors proposed in rules
{\bf iii)} and {\bf iv)}.

The tadpole cancellation condition is identical to (\ref{tadpole1}), save
for a flip in the sign of the $\delta_{k,N/2}$ contribution. They
correspond to the conditions of cancellation of the irreducible anomaly
and the RR charge conservation in the IIA brane configuration.

These two examples illustrate the construction of type IIB orientifolds
with the $\Omega$ and $\Omega \Pi$ projections. They also provide a
non-trivial check of the rules proposed for the brane configurations.
A complete classification of models is straightforward to perform, but
lenghy to list and does not provide further insights, therefore we
spare the reader their discussion.

Finally, we would like to point out that some of the
orientifolds of $\IC^2/D_K$ which can be obtained using our indications
have already appeared in section~7 of \cite{bi2}. Our discussion, however,
includes several cases not considered in this reference. We see that the
type IIA brane configurations, which inspired  the construction of our
orientifolds, have provided a really insightful guideline in our task.

\bigskip

\centerline{\bf Acknowledgements}

I am pleased to thank J.~Park for useful discussions, and the Institute
for Advanced Study, Princeton, for hospitality at the beginning of
this project. I am also grateful to M.~Gonz\'alez for her patience and
support. This work has been financially supported by the Ram\'on Areces
Foundation (Spain).

\end{document}